\documentclass[12pt]{article}
\usepackage{mathrsfs}
\def\c{\times}
\def\o{\mathrm{o}}
\def\E{\mathscr{E}}
\def\L{\mathscr{L}}

\def\RR{\mathscr{R}}

\input epsf
\usepackage{epsfig}
\def\o{\circ}

\def\const{\mathrm{const}}
\def\R{\mathbf{R}}
\def\bR{\mathbf{\overline{R}}}
\def\C{\mathbf{C}}
\def\bC{\mathbf{\overline{C}}}
\begin{document}
\title{Two-parametric PT-symmetric quartic family}
\author{Alexandre Eremenko and Andrei Gabrielov\thanks{Both authors
are supported by NSF grant DMS-1067886.}}
\maketitle
\begin{abstract}
We describe a parametrization of the real spectral
locus of the two-parametric family
of PT-symmetric quartic oscillators.
For this family, we find a parameter region where all eigenvalues are real,
extending the results of Dorey, Dunning, Tateo and Shin.


\end{abstract}

\section{A family of quartic oscillators}\label{intro}

We consider the eigenvalue problem in the complex plane
\begin{equation}\label{phys}
w^{\prime\prime}+(\zeta^4+2b\zeta^2+2iJ\zeta+\lambda)w=0,\quad w(te^{-\pi i/2\pm\pi i/3})\to 0,\;
t\to+\infty.
\end{equation}
Here $J$ and $b$ are parameters.
This two-parametric family is interesting for several reasons.

When $2J$ is an integer,
$2J<1$, and $b\geq 0$, the problem has the same
spectrum as a spherically symmetric quartic oscillator
in $\R^d$. In this case
$2J=2-2l-d$, where $l$ is the angular momentum quantum number \cite{BG,Bm}.

When $J$ is a positive integer, problem (\ref{phys}) is 
quasi-exactly solvable (QES) \cite{BB}. This means that there are $J$ eigenfunctions
of the form $$w(\zeta)=p(\zeta)\exp(-i\zeta^3/3-ib\zeta),$$ where $p$ is a polynomial
of degree $J-1$ in $\zeta$ whose coefficients are algebraic functions in $b$.

When $J$ and $b$ are real, the problem is PT-symmetric.
The eigenvalues of a PT-symmetric problem can be either real or come in complex conjugate
pairs. Both possibilities can be present for $J>1$.
A very interesting feature is level crossing in the real domain:
for some real $b$ and $J$ the graphs of the eigenvalues $\lambda_k(b)$
can be real and cross each other. This phenomenon was
discovered by Bender and Boettcher \cite{BB} numerically,
then it was studied in \cite{EG},
where the presence of infinitely many such real level crossing points was proved
for positive odd $J$.

When $J\to+\infty$, the QES part of the spectral locus approximates
the whole spectral locus of the PT-symmetric cubic family
\begin{equation}\label{cubic}
-w^{\prime\prime}+(iz^3+iaz)w=\lambda w,\quad w(\pm\infty)=0,
\end{equation}
which was subject of intensive research, see, for example,
\cite{BB2,DT,Dorey,EG0,GM,EG0,Shin,Z}.

By the change of the independent variable $z=i\zeta$ problem (\ref{phys})
is equivalent to
\begin{equation}\label{oper}
L_{b,J}(y)=-y^{\prime\prime}+(z^4-2bz^2+2Jz)y=\lambda y,\quad y(te^{\pm\pi i/3})\to 0,\quad t\to+\infty.
\end{equation}
Shin's theorem \cite{Shin} applies to these eigenvalue problems when $J\leq 0$,
and
implies that for $J\leq 0$ all eigenvalues are real. The proof of Shin's theorem is based
on the remarkable ODE-IM correspondence of Dorey, Dunning and Tateo
\cite{Dorey}.
Here we extend this result of Shin.
\vspace{.1in}

\noindent
{\bf Theorem 1.} {\em All eigenvalues of (\ref{phys}) or (\ref{oper}) are real for
$J\leq 1$.}
\vspace{.1in}

The condition $J\leq 1$ is exact, because it is known that for every $J>1$ there
are non-real eigenvalues \cite{EG}. 
Our proof of Theorem 1 is based on purely topological arguments.
Using the formulation (\ref{oper}), we establish a
certain property of eigenfunctions
for $J=0$, and then show that this property persists for $J<1$ and
prevents level crossing.

The {\em real spectral locus} $Z(\R)\subset\R^3$ is defined as
the set of all real triples $(b,J,\lambda)$
for which there exists $y\neq0$ satisfying (\ref{oper}). This is 
a non-singular analytic surface in $\R^3$.
The main result of this paper is a parametrization of a part of $Z(\R)$
corresponding to integer $J$
in terms of Nevanlinna parameters. 
In \cite{EG0} we obtained 
similar parametrization of the real spectral locus of (\ref{cubic})
and another two-parametric family of quartics.

The family considered in this paper is much more complicated
because of the presence of QES part and of the real level crossings.
In \cite{EGn} we parametrized
the real {\em quasi-exactly solvable locus} $Z^{QES}(\R)$ of (\ref{oper})
which consists of all triples $(b,J,\lambda)\in\R^3$ for which there exists
a function $y(z)=p(z)\exp(z^3/3-bz)$ satisfying (\ref{oper}) with
a polynomial $p$. 

The plan of the paper is the following.
In the next section we introduce Nevanlinna parameters and state
our principal result, Theorem 2, about the correspondence between
parameters $(b,J,\lambda)\in Z(\R)$ and Nevanlinna parameters.
In section \ref{darb} we prove the ``easy'', algebraic part of this correspondence.
In section \ref{complexes} we study the case $J=0$ and in section
\ref{J<1} we prove Theorem~1.
Then in section \ref{class-complexes} we describe the parametrization of the part
of the
real spectral locus $Z(\R)$ where $J$ is an integer in terms of Nevanlinna parameters
and complete the proof of Theorem~2.

In a forthcoming paper we will parametrize the whole two-dimensional real
spectral locus of the family (\ref{oper}) in terms of Nevanlinna
parameters.

\section{Nevanlinna parameters}\label{parameters}

The references for this section are \cite{N,Sibuya,EG-1,EG1,EG0}.

Suppose that $b$, $J$ and $\lambda$ are real. 
Let $y$ be an eigenfunction of (\ref{oper}).
Then $y^*(z)=\overline{y(\overline{z})}$ satisfies the
same differential equation and the same boundary conditions
(\ref{oper}), so
$y^*=cy$. We can choose $y$ so that $y(x_0)\in\R\backslash\{0\}$
for some
real $x_0$. Substituting
this $x_0$ to $y^*=cy$ we conclude that $c=1$. So our eigenfunction $y$ is real.
It is defined up to multiplication by a real constant.

Let $y_1$ be a solution of
the differential equation
in (\ref{oper}), which is real and linearly independent of $y$.
Consider the meromorphic function $f=y/y_1$. It is a {\em Nevanlinna function},
which means that $f$ has no critical points in $\C$ and the only singularities
of $f^{-1}$ are finitely many logarithmic branch points.

In the case
that $y_1$ is normalized by $y_1(x)\to 0,\; x\to+\infty,\; x\in\R$,
we call $f$ {\em a normalized} Nevanlinna
function of (\ref{oper}). The normalized Nevanlinna function is defined up to
a real multiple. Existence of $y_1$ with such normalization is guaranteed
by a theorem of Sibuya \cite{Sibuya}.

Different Nevanlinna functions associated with the same point
$(b,J,\lambda)\in Z(\R)$ are related by a real fractional-linear 
transformation of the form $f\mapsto \alpha f/(f-\beta)$
where $\alpha\neq 0$ and $\beta$ are real.

Nevanlinna functions $f$ of (\ref{oper}) have no critical points,
because $f'=(y'y_1-yy_1^\prime)/y_1^2$ and 
$y'y_1-yy_1^\prime=\const$. They have $6$ asymptotic values in the sectors
$$S_j=\{ te^{i\theta}: t>0,\; |\theta-\pi j/3|<\pi/6\},\quad j=0,\ldots,5.$$
In what follows $j$ is understood as a residue modulo $6$.
These sectors are in one-to-one correspondence with logarithmic branch points
of $f^{-1}$.
The asymptotic values of the normalized Nevanlinna function
are $0$ in $S_1$ and $S_{-1}$, because of the boundary condition, and 
$\infty$ in $S_0$ because of the normalization of $y_1$.
We denote by $c$ and $a$ the asymptotic values
of the normalized Nevanlinna function in $S_2$ and $S_{3}$,
respectively. As $f$ is real, the asymptotic value in $S_{-2}$ is $\overline{c}$,
and $a$ is real. 

These asymptotic values $a,c$ are called the {\em Nevanlinna parameters}.
They are related to the Stokes multipliers by simple formulas
\cite{Sibuya,Masoero}. Normalized function $f$
and the Nevanlinna parameters are
defined modulo multiplication by a real non-zero number, so we can
further normalize them. We will use different normalizations, depending
on the situation.

Nevanlinna parameters $(c,a)\in\bC\times(\R\cup\{\infty\})$
are subject to the following conditions:
\begin{equation}\label{conditions}
c\neq 0,\quad c\neq a,
\end{equation}
and the set $\{0,\infty,a,c,\overline{c}\}$ contains at least $3$ distinct
points, in other words, the combination $c=\infty,\; a=0$ is prohibited
\cite{N,Sibuya}.

Nevanlinna parameters (modulo multiplication by real constants)
serve as local coordinates on the real spectral locus $Z(\R)$,
\cite{Bakken,Masoero}. Notice that the set of pairs satisfying (\ref{conditions})
modulo proportionality is a non-Hausdorff manifold.
But this will cause no difficulties as we always work in local charts.

Relation between $(b,J,\lambda)$ and $(a,c)$ is very complicated:
Nevanlinna's construction of the map $(a,c)\to(b,J,\lambda)$ involves the
uniformization theorem. So it is interesting and challenging
to establish any explicit
correspondences between the sets in the space of Nevanlinna parameters
and the sets in the $(b,J,\lambda)$ space.

Our main result in this direction is
\vspace{.1in}

\noindent
{\bf Theorem 2.} {\em $J$ is an integer if and only if 
$$a=0\quad\mbox{or}\quad c=\overline{c}.$$}

In the next section we prove the ``only if'' part.

\section{QES locus and Darboux transform}\label{darb}

In this section we prove the ``easy part'' of Theorem 2:
if $J$ is an integer, then either $a=0$ or $c=\overline{c}$.

Suppose that for some $(b,J,\lambda)\in Z(\R)$ we
have $a=0$. This means that the eigenfunction $y$ tends to zero in $S_3$
(and also in $S_1,S_{-1}$), so $y$ is an elementary function of the form
$pe^q$ with polynomials $p$ and $q$. Substitution to (\ref{oper}) gives
$$y(z)=p(z)\exp(z^3/3-bz),$$
where $p$ is a polynomial. It is known that such eigenfunctions exist
if and only if $J$ is a positive integer \cite{BB,EG}. Such points
$(b,J,\lambda)$ form the QES spectral locus $Z^{QES}(\R)$ which  
consists of smooth algebraic curves
$$Q_J(b,\lambda)=0,\quad J=n+1,\quad n\geq 0,$$
where $Q_J$ are real polynomials of degree $J$ in $\lambda$.
QES spectral locus was studied in \cite{EG,EGn}; in the second paper
it was parametrized in terms of Nevanlinna parameters.
The converse is evident: if $(b,J,\lambda)\in Z^{QES}(\R)$, then $a=0$.

In \cite{EG} we obtained the following results:
\vspace{.1in}

\noindent
{\em (i) If $J$ is a positive integer, and $b$ is real, then all non-QES eigenvalues are
real.}
\vspace{.1in}

Let $Z_{J}(\R)=\{(b,\lambda)\in\R^2:(b,J,\lambda)\in Z(\R)\}$
and let $Z_{J}^{QES}(\R)$
be similarly defined. Let $Z_{J}^*$ be the closure 
of
$Z_{J}(\R)\backslash Z_{J}^{QES}(\R)$ in $\R^2$.
\vspace{.1in}

\noindent
{\em (ii) When $J$ is even, $Z_{J}^*(\R)\cap Z_{J}^{QES}(\R)=\emptyset.$
When $J$ is odd,
then $c=\overline{c}$ holds at all points $(b,\lambda)$ of this intersection.
\vspace{.1in}

\noindent
(iii) If $J$ is a positive integer, then $Z_J^*=Z_{-J}$.}
\vspace{.1in}

To deal with the condition $c=\overline{c}$, we use the Darboux
transform \cite{Crum,Veselov,EG}, which we recall.
Let $\psi_0,\ldots,\psi_n$ be some eigenfunctions of a differential
operator $L=-d^2/dz^2+V(z)$ with eigenvalues $\lambda_0,\ldots,\lambda_n$.
Then the differential operator
\begin{equation}\label{darboux}
-\frac{d^2}{dz^2}+V-2\left(\log W(\psi_0,\ldots,\psi_n)\right)^{\prime\prime},
\end{equation}
where $W$ is the Wronski determinant, has the same eigenvalues as $L$,
{\em except} $\lambda_0,\ldots,\lambda_n$.

Let $J\geq 0$ be an integer, $(b,\lambda)\in Z_J^*(\R)$,
and let $y$ be the eigenfunction corresponding to $(b,J,\lambda)$.
We apply the Darboux transform to our operator $L_{b,J}$ in (\ref{oper}),
taking all QES eigenfunctions as $\psi_0,\ldots,\psi_n$.
If $J=0$ the Darboux transform does not change anything.
It is easy to see, \cite{EG} that the transformed operator (\ref{darboux})
is $L_{b,-J}$ in this case. But $L_{b,J}(y)=\lambda y$, and if we define
$y^*(z)=y(-z)$, then $L_{b,-J}(y^*)=\lambda y^*.$
However, this $y^*$ is not an eigenfunction of $L_{b,-J}$, because
it does not satisfy the normalization condition in (\ref{oper}).
Instead it tends to zero in $S_2$ and in $S_{-2}$. This means that
$y^*$ is linearly independent of the eigenfunction $y_0$ of $L_{b,-J}$,
and $g=y_0/y^*$ has asymptotic values $\infty$ in $S_2$ and in $S_{-2}$.

Since $g$ has equal asymptotic values in $S_2$ and $S_{-2}$, any other Nevanlinna
function for the same $(b,J,\lambda)$ has the same property.
So
$c=\overline{c}$ at the point $(b,-J,\lambda)$.

The case of negative $J$ is treated similarly, applying the inverse Darboux
transform. Thus $c=\overline{c}$ on $Z_J^*(\R)$
when $J$ is an integer. This proves the ``only if'' part of Theorem~2.

As a byproduct we obtain the following:
\vspace{.1in}

\noindent
{\em (iv) 
condition
$c=\infty$ at the point $(b,\lambda)\in Z_{-J}(\R)$ holds if and only if
$J>0$ is odd and
$(b,\lambda)\in Z_J^{QES}(\R)\cap Z_J^*(\R)$} 
\vspace{.1in}

Indeed, the eigenfunction $y$ is elementary if and only if 
the asymptotic values of $f$ in $S_3,S_1,S_{-1}$ are zero.
This happens if and only if the function $g$ defined above has asymptotic values
$\infty$ in  $S_0,S_2,S_{-2}$.  

To prove the second part of Theorem 2,
we need to find which part of the spectral locus
corresponds to real $c$.

\section{Line complexes}\label{complexes}

Assigning asymptotic values in sectors $S_j$ is not enough
to define a Nevanlinna function $f$, one needs additional information
about the topology of the covering
$$f:\C\backslash f^{-1}(\mbox{asymptotic values})\to
\bC\backslash\{\mbox{asymptotic values}\}.$$
Such information is encoded in the following way \cite{N,EG-1,EG1}.
One chooses a cell
decomposition $\Phi$ of the Riemann sphere $\bC$
such that each 2-cell contains
one asymptotic value, and takes the $f$-preimage of this cell decomposition.
This preimage is a cell decomposition $\Psi$ of the plane which locally looks
like $\Phi$. There are many ways to choose $\Phi$, and here we describe
cell decompositions used by Nevanlinna, see also \cite[Ch. 7]{GO} for a
comprehensive treatment. 

First we fix two points in $\bC$ which are distinct from the asymptotic values.
We call these points $\c$ and $\o$. They are the vertices of $\Phi$.
Suppose that we have $q$ asymptotic values. We connect $\c$ and $\o$ with
$q$ edges which do not intersect except at the ends. These edges
and vertices form the $1$-skeleton of $\Phi$; we require that
faces of $\Phi$ contain one asymptotic value each.
This choice of $\Phi$ is fixed in this section.

The preimage $\Psi=f^{-1}(\Phi)$ is called the {\em line complex}.
It has the following property: {\em the $1$-skeleton of $\Psi$ is a bipartite
connected graph embedded in $\C$ whose all vertices have degree $q$,
and each component of the complement has either two or infinitely
many edges on the boundary. Moreover, the number of components
having infinitely many boundary edges is finite.}
These properties completely characterize all possible line complexes
arising from Nevanlinna functions.

This means that for arbitrary line complex,
and asymp\-totic values (Na\-evan\-lin\-na pa\-ra\-me\-ters), 
satisfying the res\-trictions stated in section~\ref{parameters}, there exists
a normalized Ne\-van\-lin\-na function with this line complex and these asymp\-totic
values, 
and it is unique up to an affine change
of the independent
variable. The Schwarzian derivative of this function
is a polynomial and we compose with an affine change of the independent
variable so that this polynomial has leading coefficient $-2$
and next coefficient vanishing, so that
\begin{equation}\label{schwarz}
\frac{f^{\prime\prime\prime}}{f^\prime}-
\frac{3}{2}\left(\frac{f^{\prime\prime}}{f^\prime}\right)^2=
-2(z^{d}+a_{d-2}z^{d-2}+\ldots+a_0).
\end{equation}
In this paper, $d=4$, $(a_2,a_1,a_0)=(2b,2iJ,\lambda)$ and Nevanlinna
parameters are asymptotic values $a$ and $c$ as described in
section~\ref{parameters}. Function $f$ does not change if $a$ and $c$ are multiplied
by the same number.
For a fixed line complex, the map that sends the 
parameter $t=c/a$ to the Nevanlinna triple $(b,J,\lambda)$ is called the 
{\em Nevanlinna map}. It is smooth, injective and has
non-zero derivative. It gives a parametrization
of a part of the spectral locus. 

We label the faces of $\Phi$ and $\Psi$ with the asymptotic values,
so that a face of $\Psi$ has the same label as its image.
Two cell decompositions $\Psi^\prime$ and $\Psi^{\prime\prime}$ are
considered equivalent if one can be mapped onto another
by a homeomorphism of the plane preserving orientation and labels of
faces and vertices. If two Nevanlinna functions $f$ and $g$ have equivalent
cell decompositions, then $f(z)=g(\alpha z+\beta),\;\alpha\neq 0$.

The following properties are evident. The cyclic order of face labels
around a vertex of $\Psi$ is the same as for the image vertex in $\Phi$.
Two faces of $\Psi$ with the same labels have disjoint closures.
Each face is bounded either by two edges of by infinitely many edges.

By erasing multiple edges of the $1$-skeleton of $\Psi$ and
discarding the labels of bounded faces, we obtain a new cell decomposition
with labeled faces and vertices, whose $1$-skeleton is a tree.
The line complex can be uniquely recovered from its associated tree.

If $f$ is real, so asymptotic values are symmetric with respect to complex
conjugation, sometimes it is possible to choose a symmetric cell
decomposition $\Phi$. Then $\Psi$ is also symmetric. And conversely:
if the labeled cell decompositions
$\Phi$ and $\Psi$ are symmetric with respect to complex conjugation
then $f$ can be chosen real by pre-composing with an affine map of $\C$.

Line complexes are convenient for study of the limits of families
of Nevanlinna functions when two
asymptotic values collide.
We have the following compactness theorem \cite{Volk}.
Fix a cell decomposition $\Phi$. Let $f_n$ be a sequence
of Nevanlinna functions with line complexes $\Psi_n=f_n^{-1}(\Phi)$.
Suppose that $v_n=0$ is a vertex of $\Psi_n$, and that
$f_n$ are normalized by conditions $|f_n^\prime(0)|=1$.
Then one can choose a subsequence from $f_n$ that tends to a limit,
and this limit is a Nevanlinna function.

We need only the special case when the cell decompositions $\Psi_n$
corresponding to $f_n$ are all equivalent.
If distinct asymptotic values of $f_n$
tend to distinct limits, then $f$ has the  same cell decomposition
$\Psi$.
If two asymptotic values of $f_n$ which are labeling adjacent faces of $\Phi$,
collide in the limit, then one has to
erase from the $1$-skeleton of $\Psi$ all edges on the common boundaries
of faces with these collided asymptotic values.
The component of the remaining graph containing the vertex $v$ is the cell
decomposition of the limit function.

Now we return to Nevanlinna functions corresponding to problem
(\ref{oper}). One technical problem we are facing is that
it is not always possible to choose a symmetric $\Phi$.
However this is possible when $c$ is real, and in the next sections we consider this case.
We begin with the simplest case when $J=0$. In this case the Nevanlinna
function has an additional symmetry.

The cell decompositions considered in \cite{EGn} are {\em different}
from line complexes, because in the situation considered in that paper,
it is impossible to define a line complex with the required symmetry
properties.

\section{Subfamily $J=0$}\label{J=0}

In this section we begin to prove Theorem 1.

Let $y$ be an eigenfunction  of $L_{b,0}$, where $b\in\C$. 
Function $y_1(z)=y(-z)$ satisfies the differential
equation in (\ref{oper}) with $J=0$,
but does not satisfy the boundary conditions.
So $y_1$ is linearly independent of $y$, and we consider
the Nevanlinna function
\begin{equation}\label{f}
f=y/y_1.
\end{equation}
It is {\em not normalized} in the sense of section \ref{parameters}.
This function $f$ has the following symmetry property:
\begin{equation}\label{condition}
f(-z)=1/f(z).
\end{equation}
The asymptotic values are $0$ in $S_1$ and $S_{-1}$,
$\infty$ in $S_2$ and $S_{-2}$,
$A$ in $S_0$ and $1/A$ in $S_3$. This $A$ is the Nevanlinna parameter.
It follows from (\ref{schwarz}) that the condition (\ref{condition})
is equivalent to $J=0$.

If $b$ is real, we know from the results of \cite{Bm} and
\cite{Shin} that all eigenvalues of $L_{0,b}$ are real.
Hence $f$ is a real function and $A\in\R$.
As $f$ is defined up to multiplication by a real non-zero number,
we can normalize
so that $A>0$. 
\vspace{.1in}

\noindent
{\bf Proposition 1.} {\em For $b\in\R$, we have $A\in(0,1)$. 
Each eigenfunction has at most one zero on the real line.}
\vspace{.1in}

{\em Proof.} Function $f$ has the property that
$$f:\C\backslash f^{-1}(\{0,A,1/A,\infty\})\to\bC\backslash\{0,A,1/A,\infty\}$$
is a covering. To construct the line complex,
we choose the cell decomposition $\Phi$ of the
target sphere shown in Fig.~1.

\bigskip
\begin{center}
\epsfxsize=2.0in%
\centerline{\epsffile{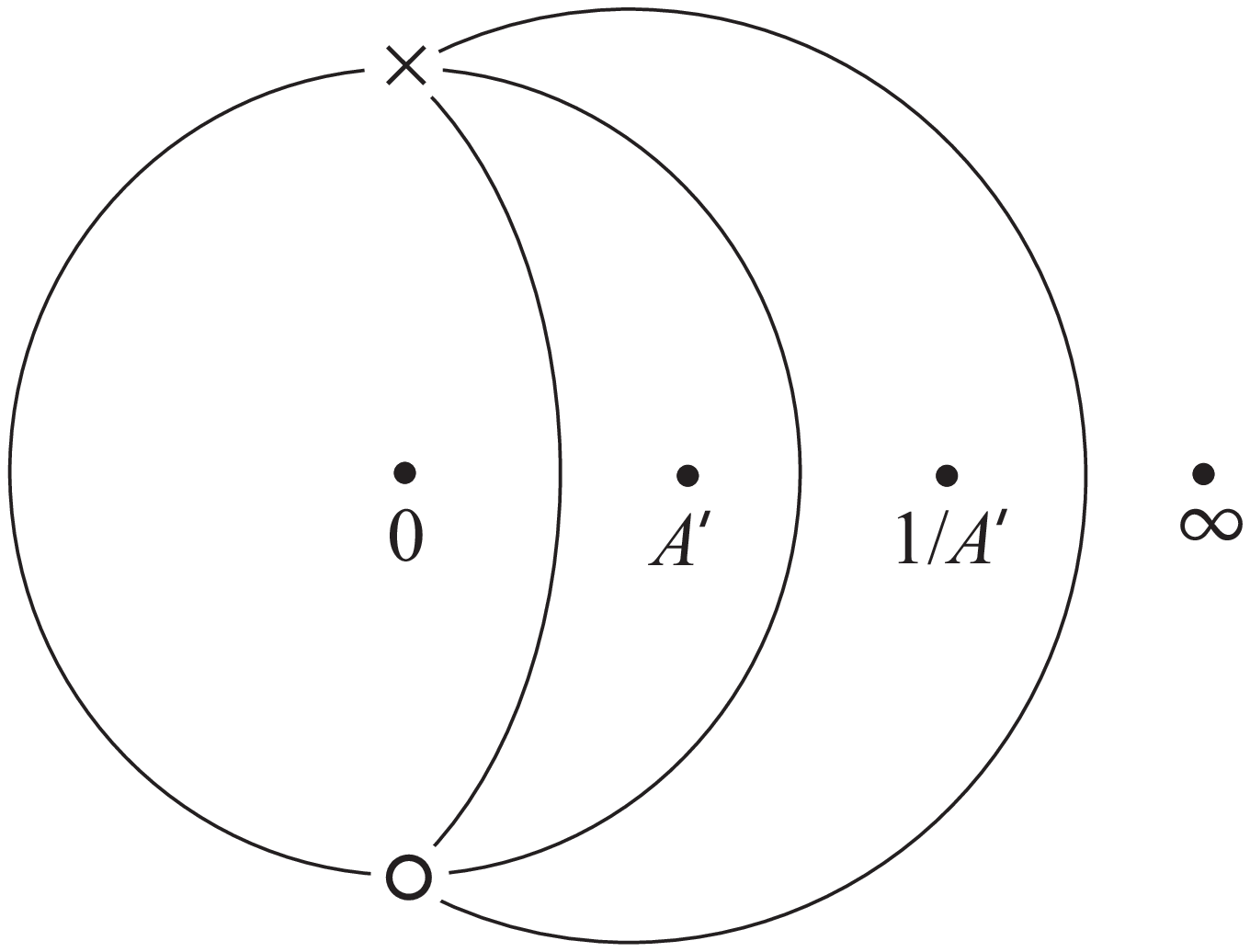}}
\nopagebreak\medskip
\noindent Fig. 1. Cell decomposition $\Phi$ of the Riemann sphere. 
\end{center}
\bigskip

It has two vertices, 
four edges, and four faces labeled by the asymptotic values.
We denote by $A'$
the asymptotic value which is in $(0,1)$,
so that $A'$ is either $A$ or $A^{-1}$,
and our first goal is to find out which of these possibilities holds.

The line complex $\Psi=f^{-1}(\Phi)$
is a labeled cell decomposition of $\C$.
It has $6$ unbounded faces.
Moreover, $\Psi$ is symmetric with respect to the real line
and with respect to
the imaginary line. The symmetry with respect to the
real line does not change
the labels, while the symmetry with respect to the imaginary
line interchanges $\c$ with
$\o$, $0$ with $\infty$ and $A'$ with $1/A^\prime$.
Unbounded faces of $\Psi$
are asymptotic to the sectors $S_j$.

For any pair of vertices of $\Psi$ connected by several edges,
we replace
these several edges with one edge.
The result is a simpler cell decomposition $T$
whose $1$-skeleton is a tree.
The faces of $T$ are labeled with asymptotic values,
and $T$ has all symmetry properties described above.
The label of a face asymptotic
to $S_j$ is the asymptotic value in $S_j$.

It is easy to classify all possible labeled trees
satisfying the above conditions. They all have two vertices
of order 4 and the number $k$ of edges between these two vertices
is odd.
These trees depend on one non-negative integer parameter $m$, such that $k=2m+1$,
and we denote them by $T_m$.
The tree $T_1$ and the corresponding line complex are shown
in Fig.~2.
\bigskip

\begin{center}
\epsfxsize=4.0in%
\centerline{\epsffile{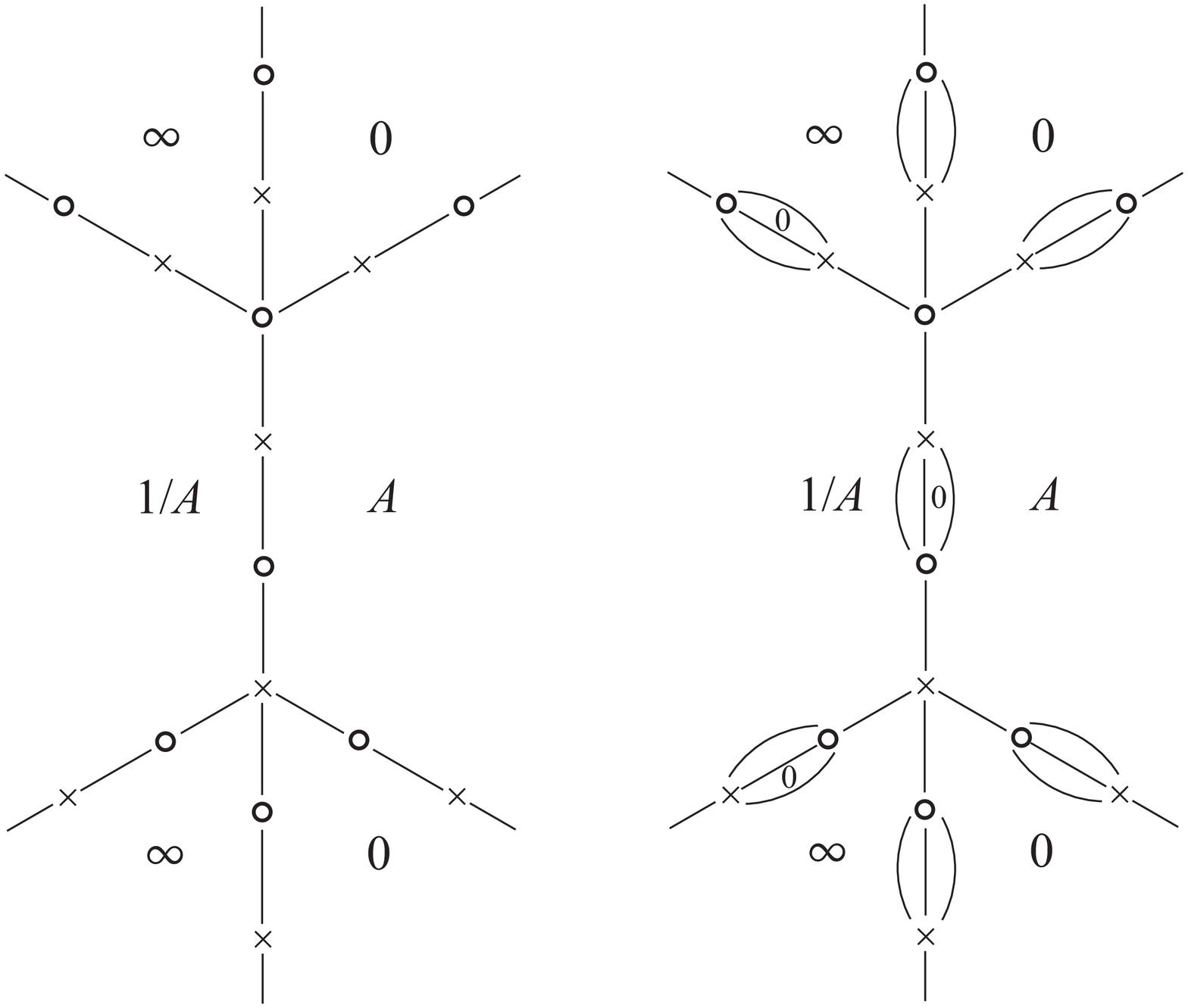}}
\nopagebreak\medskip
\noindent Fig. 2. The tree $T_1$ (left) and the corresponding line complex (right).
\end{center}
\bigskip

Comparing the cyclic order of
labels
of faces adjacent to a vertex of order $4$ in Fig.~1 and Fig.~2 we conclude that
$A=A'$ so $A\in(0,1)$. This proves the first part of the proposition.
The second part is immediately clear from the classification of
the trees $T_m$: the cell decomposition $\Psi$ has at most one face labeled with $0$
which intersects the real line. (A similar argument was used in \cite{EG-1}.)
Notice that
\begin{equation}\label{eo}
\mbox{the number of real zeros of $y$ is $0$ if $m$ is even and $1$ if $m$ is odd.}
\end{equation}

\noindent
{\bf Lemma 1.} {\em For $b\in\R$, eigenvalues of $L_{0,b}$ are distinct.}
\vspace{.1in}

{\em Proof.}
The subset of the spectral locus $Z$ parametrized
by the Nevanlinna functions satisfying (\ref{condition}) 
coincides with $Z_0$.
Since Nevanlinna parameters (modulo multiplication by non-zero numbers)
serve as a local coordinate system on $Z$,
$Z_0$ is a smooth one-dimensional subset of $Z$.
Hence $Z_0(\R)$ consists of smooth real curves and, possibly,
isolated points. Since all eigenvalues of $L_{0,b}$ are real
for $b\in\R$, isolated points are not allowed.
Hence $Z_0(\R)$ is smooth, and the real eigenvalues 
of $L_{b,0}$ cannot collide as $b\in\R$ changes. 
This proves the lemma.
\vspace{.1in}

Thus we can label the eigenvalues in increasing order,
$\lambda_0(b)<\lambda_1(b)<\ldots$.
To each eigenvalue $\lambda_k$ corresponds
one cell decomposition $\Psi$, and one
tree $T=T_m$.
So we have some correspondence $m(k)$, which is defined so that $T_{m(k)}$
corresponds to $\lambda_k$.
\vspace{.1in}

\noindent
{\bf Proposition 2.} {\em $m(k)=k.$ So the eigenfunction $y_k$ corresponding to $\lambda_k$
has no real zeros when $k$ is even and one real zero when $k$ is odd.}
\vspace{.1in}

{\em Proof.} We use the asymptotic result in \cite{EG,EG0} to degenerate (\ref{oper})
as $b\to+\infty$ to the harmonic oscillator
$$Y^{\prime\prime}(u)+4u^2Y(u)=\mu Y(u),\quad Y(it)\to0,\quad t\to\pm\infty.$$
The boundary condition here comes from the boundary 
condition in (\ref{oper}). 
Eigenfunctions $Y_k$ corresponding to eigenvalues $\mu_k$, labeled in the
increasing order, have all zeros on the imaginary axis.
There is one zero on the real axis (namely at the origin) when $k$ is odd and none
if $k$ is even.

To analyse the behavior of zeros of eigenfunctions of (\ref{oper}) as $b\to+\infty$,
we consider the function
$g=f/(f-A)$. This corresponds to a different choice of $y_1$ in the basis $(y,y_1)$
of solutions of the differential equation in (\ref{oper}).
Asymptotic values of $g$ are $\infty$ in $S_0$, $0$ in $S_{\pm1}$, $1$ in $S_{\pm2}$,
and $a:=1/(1-A^2)$ in $S_3$.

According to the result of \cite{EG0}, to every eigenfunction $Y_k$
with eigenvalue $\mu_k$, and to every positive $b$ large enough,
corresponds a unique eigenfunction $y$ of (\ref{oper}) with eigenvalue $\lambda(b)\to\mu_k$.
Let $m=m(k)$ and let $T_m$ be the tree corresponding to $y$.
Then $T_m$ must collapse to the tree $T^\prime_k$ corresponding to $Y_k$
(see Fig.~3).
\bigskip

\begin{center}
\epsfxsize=1.5in%
\centerline{\epsffile{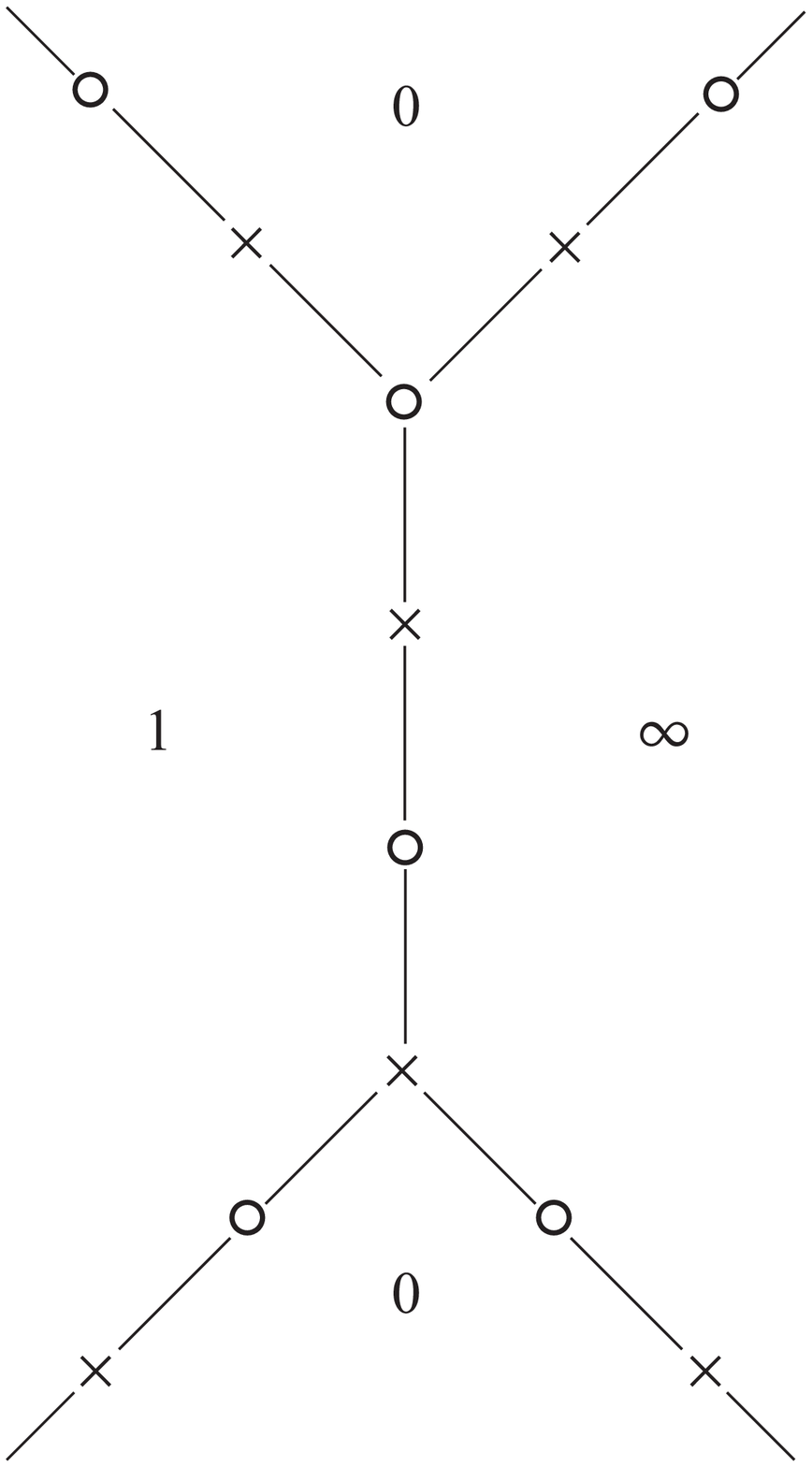}}
\nopagebreak\medskip
\noindent Fig. 3. Tree $T^\prime_1$.
\end{center}
\bigskip

By collapse of a tree we mean the following: two face labels become equal, and all
edges on the common boundary of these faces are erased.
Such collapse can happen only if $1/(1-A^2)\to 1$ or
$1/(1-A^2)\to\infty$. In the second case we must multiply $g$
by $1-A^2$, so the asymptotic value in $S_3$ becomes $1$ and in $S_{\pm2}$ it becomes 
$1-A^2\to 0$.
Then as the tree $T_m(k)$ collapses to the tree $T^\prime_k$, we must have
$m(k)=k$. This proves the proposition. 
\vspace{.1in}

Proposition 2 gives a parametrization of the spectral locus for $J=0$
by Nevanlinna
parameter $A$. For each $m$, the correspondence $b\to A$ is a real analytic
homeomorphism of the real line onto $(0,1)$. 

\section{Region $J<1$}\label{J<1}

Now we can prove Theorem 1.
When $b,J$ and $\lambda$ are real, the eigenvalue problem is PT-symmetric, and we can define
a real eigenfunction $y$. We choose the second linearly independent solution $y_1$
of the differential equation in (\ref{oper}) from the conditions that $y_1\to 0$
in $S_0$, and $y_1$ is real. Then $f=y/y_1$ is real. The asymptotic values of $f$
in $S_0,\ldots,S_5$ are $\infty,0,c,a,\overline{c},0$ in this order, where
$a\in\R$. Thus for every
$(b,J,\lambda)$ on the real spectral locus, Nevanlinna parameters $a\in\bR$ and $c\in\bC$
are defined.

As long as $a\neq 0$, the number of real zeros cannot change, because $y$ and $f$ never have
multiple zeros.

Equality $a=0$ can happen on the spectral locus if and only if
$J$ is a positive integer and the eigenfunction $y$ is elementary \cite{BB,EG}.
\vspace{.1in}

\noindent
{\bf Lemma 2.} {\em Let $\gamma(t), t\in[0,1]$ be a curve in
the real $(b,J)$ plane.
Suppose that $\gamma(0)=(0,0)$, and
that $J\leq 1$ on $\gamma$.
Then for $(b,J)=\gamma(1)$, all eigenvalues are real.}
\vspace{.1in}

\noindent
This implies Theorem~1. 
\vspace{.1in}

{\em Proof of Lemma 2.} It is sufficient to prove the lemma for curves in
the open half-plane $J<1$. The general case follows by continuity.

For $(b,J)=(0,0)$ all eigenvalues are real and distinct, so we
can
order them as $\lambda_0<\lambda_1<\ldots.$ Then each $\lambda_k$ can be analytically
continued along $\gamma$ for $t\in[0,t_k)$ for some $t_k>0$. We denote these
analytic continuations by $\lambda_k(t)$. According to theorems of Shin
\cite{Shin0} all but finitely many of the $\lambda_k$ are real 
and by \cite[Thm. 1]{EG1}, all but finitely many $\lambda_k$ can be analytically continued
for $t\in[0,1]$. Let $G$ be a bounded simply connected
neighborhood of $\gamma$ in the half-plane $\{(b,J)\in\R^2:J<1\}$.
Then all but finitely many branches of $\lambda(b,J)$ are holomorphic, distinct
and real in $G$. The remaining branches satisfy a minimal algebraic equation
of the form
\begin{equation}\label{eq}
\lambda^m+a_{m-1}(b,J)\lambda^{m-1}+\ldots+a_m(b,J)=0,
\end{equation}
with $a_j$ analytic in $G$.
The zeros of the discriminant of this equation in $G$ form a closed subset $K\subset G$.

We claim that all solutions of equation (\ref{eq}) are distinct for all
$(b,J)\in\gamma$. Indeed, if any two eigenvalues collide 
as $t$ increases, then some adjacent eigenvalues $\lambda_j(t)$ and $\lambda_{j+1}(t)$
must collide
at some point $t_0$.
The corresponding eigenfunctions will tend to the same limit as $t\nearrow t_0$.
But this is
impossible because one of them has no real zeros and another has one. 

Thus $K$ does not intersect $\gamma$, and there is an analytic
continuation of eigenfunctions to $\gamma(1)$.

\section{Classification of line complexes}\label{class-complexes}

In this section we consider the general case of real
asymptotic values $a$ and $c$. In this case the normalized Nevanlinna function $f$
has asymptotic values $(\infty,0,c,a,c,0)$ in $S_0,\ldots,S_5$.
We choose the cell decomposition $\Phi$ similar to that in Fig.~1, see Figs.~5,7,9.

There are three generic cases:
$$\mbox{Case}\; \L.\quad c<0<a, \quad\mbox{Case}\; \RR.\quad
0<a<c,\quad\mbox{Case} \;\E.\quad 0<c<a.$$
Letters $\L,\,\RR$ and $\E$ stand for ``left'', ``right'' and
``even'', the meaning
of this notation 
will be clear later (Figs.~11-15). 
There are also non-generic cases $a=0$ and $c=\infty$.
Assuming that $a\neq 0$, we classify all possible line complexes.

In all cases $\L,\;\RR$ and $\E$, we first classify all possible bipartite trees symmetric
with respect to the real line, with $6$ faces 
labeled by $\infty,0,c,a,c,0$ in this cyclic order, 
the face labeled $\infty$ bisected by the positive ray,
and satisfying the condition that
faces with the same label have disjoint closures.
There are three types of such trees shown in Fig.~4. They depend on
two integer parameters, $k>0$ and $l$,
where $k$ is the number of edges between 
two ramified vertices as shown in Fig.~4.
Parameter $l$ takes all integer values, but $|l|$ is the number of
edges between ramified vertices as indicated in Fig.~4.
We say that a tree has type $0$ if $l=0$,
type $1$ if $l>0$ and type $2$ if $l<0$.

For the trees
that occured in section \ref{J=0} we have $T_n=X_{2n+1,0}.$
\bigskip

\begin{center}
\epsfxsize=5.0in%
\centerline{\epsffile{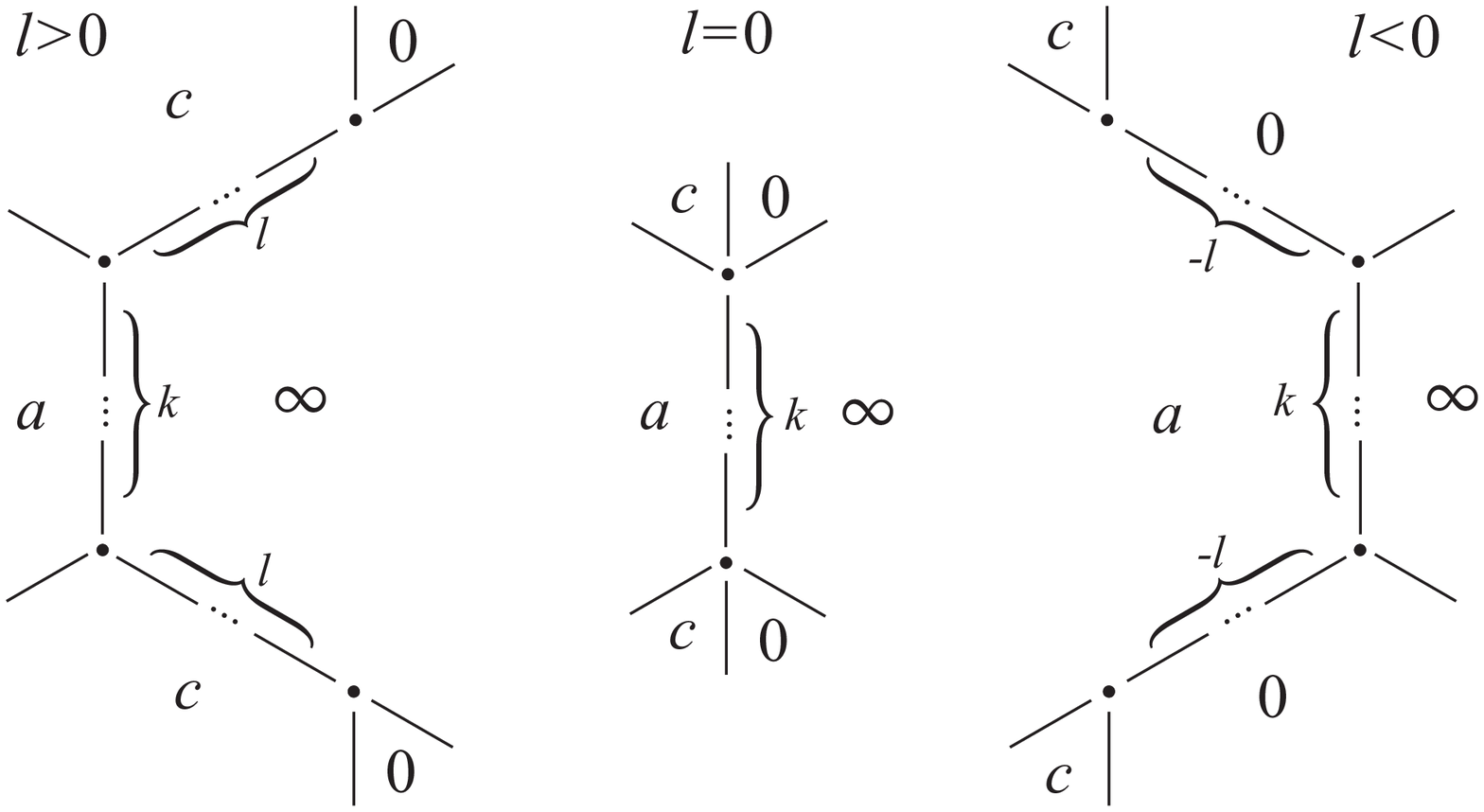}}
\nopagebreak\medskip
\noindent Fig. 4. Three types of trees.
\end{center}
\bigskip

Now we consider all cases separately, and argue by the following scheme.
First, for a given case of ordering $a,c,0$ on the real line,
and for each tree, we decide whether this tree can come from a line complex, and
if it can, we recover
the line complex.

To do this, we begin with a ramified vertex $v$ of the tree.
Comparing the cyclic order of the labels around this vertex $v$
with the cyclic order around the vertices
of the cell decomposition $\Phi$
of the sphere we determine whether this vertex of the tree is a
$\c$ or $\o$, and add the missing edges, step by step.
When we come to another ramified vertex, either the cyclic order
is correct or not. If it is not correct, the tree does not correspond
to a line complex in the considered case. Otherwise, we recover
the line complex uniquely.

We write the Nevanlinna map defined in section~\ref{complexes}
as $(b,J,\lambda)=F(\Psi,t),$ where $\Psi$ is the line complex, and $t=c/a$.

Then we consider possible limits as $a\to 0$ or $c\to\infty$ on
the spectral locus. This gives degenerations to the intersections
of the QES spectral locus with the non-QES spectral locus, and description
of these intersections in \cite{EGn}
permits to recover the value of $J$ from the tree in cases 1 and 2.
Then we consider the degeneration as $c\to\infty$ or $c\to a$
whenever possible, as it was done in \cite{EG0,EG,EGn}.
We will conclude that real values of $c$ correspond
to integer $J$ and we obtain the parametrization of the whole
spectral locus for integer $J$.
\bigskip

\noindent
{\em Case $\L$.} $\quad c<0<a$, see Figs.~5--6.
\bigskip

\begin{center}
\epsfxsize=2.0in%
\centerline{\epsffile{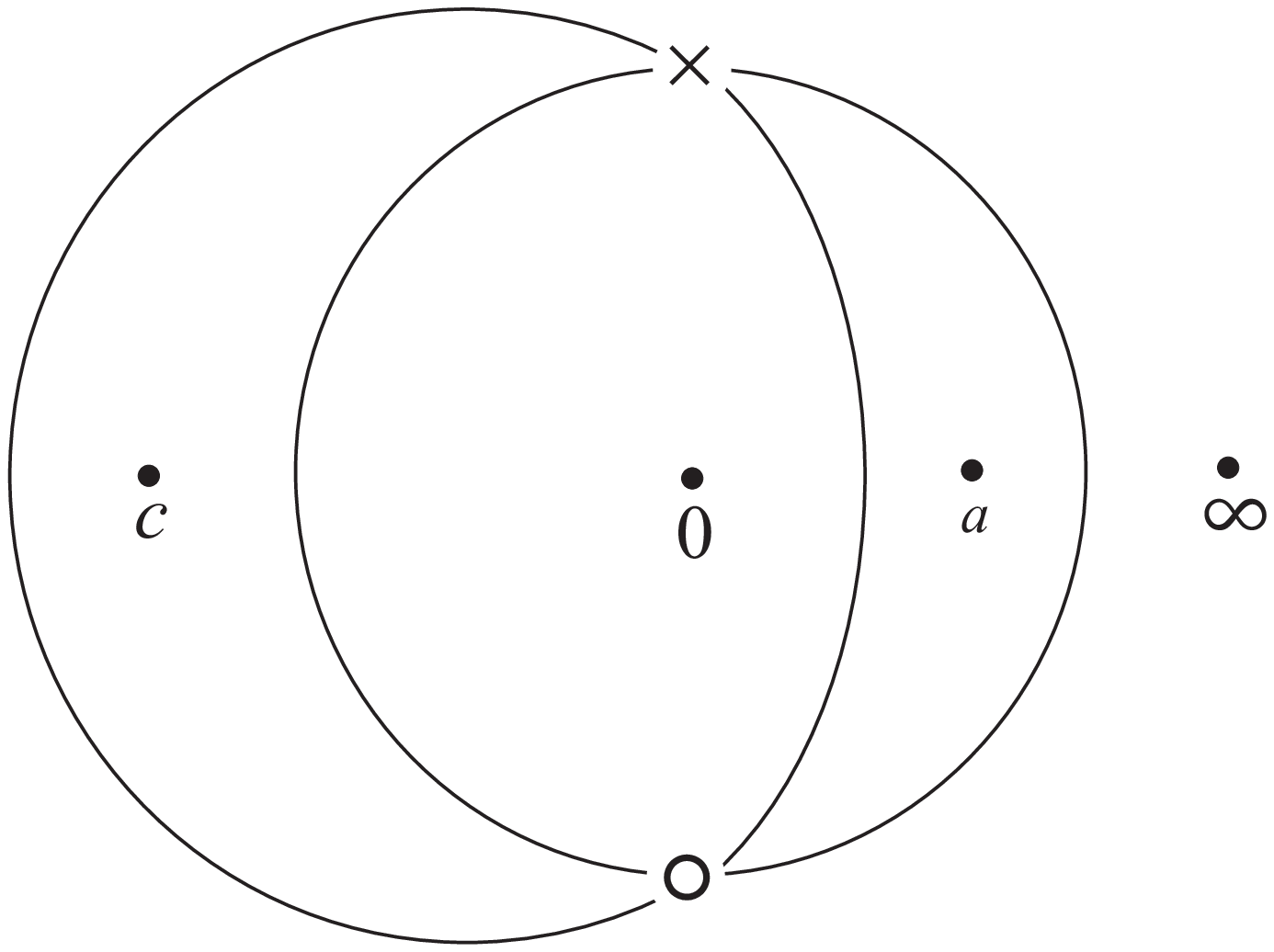}}
\nopagebreak\medskip
\noindent Fig. 5. Cell decomposition $\Phi$ in case $\L$.
\end{center}

\bigskip

\begin{center}
\epsfxsize=4.0in%
\centerline{\epsffile{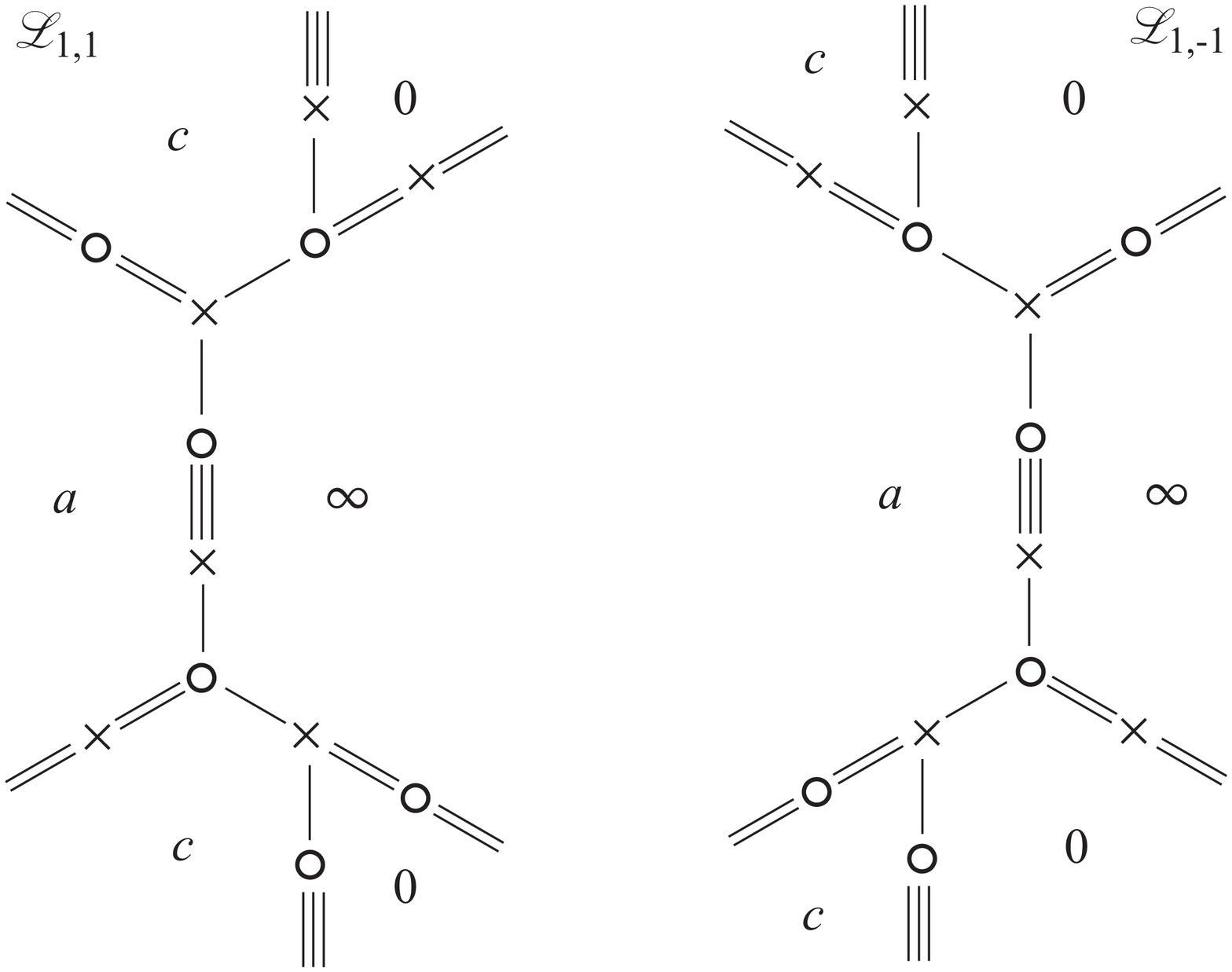}}
\nopagebreak\medskip
\noindent Fig. 6. $\L_{m,l}$ complexes.
\end{center}
\bigskip

Trees of type $0$ (with $l=0$) are impossible in this case.

Trees $X_{k,l}$ of type $1$ (with $l>0$) are possible in
this case if and only if $k=2m+1,\; m=0,1,\ldots$ and $l\geq 1$ is odd.
Such line complexes will be called $\L_{m,l}$. When $a\searrow 0$,
$t=c/a\to-\infty$,
the corresponding Nevanlinna function has a limit on the QES locus.
This limit function has $l-1$ zeros, none of them real. It follows that
$J=l$ for the limit function. By the first part of Theorem 2 which we proved in
section \ref{darboux}, $c$ is real for integer $J$ on the non-QES part of
the spectral locus, so we conclude that the whole image of the
Nevanlinna map $t\mapsto F(\L_{m,l},t)$ belongs to $Z_J(\R)$ with $J=l$.

In the limit when $c\nearrow 0$, that is $t=c/a\nearrow 0$,
we obtain a Nevanlinna function
for the harmonic oscillator with $m$ zeros. 

Trees of type $2$ (with $l<0$) are possible in this case if and only if 
$k=2m+1$ and $l\leq -1$ is odd. Such line complexes will be called
$\L_{m,l}$. The only possible limit on the spectral locus is
$c\to-\infty$. This corresponds to an elementary {\em second} solution
of the differential equation in (\ref{oper}) (solution which is linearly
independent of the eigenfunction). These points are marked in Fig.~15.
Thus the chart $\L_{m,l}$, $l\leq -1$ corresponds to the chart $\L_{m,-l}$
via Darboux transform. We have $J=l<0$ in this case.
\bigskip

\noindent
{\em Case $\RR$.} $\quad 0<a<c$, Figs.~7--8.
\bigskip

\begin{center}
\epsfxsize=2.0in%
\centerline{\epsffile{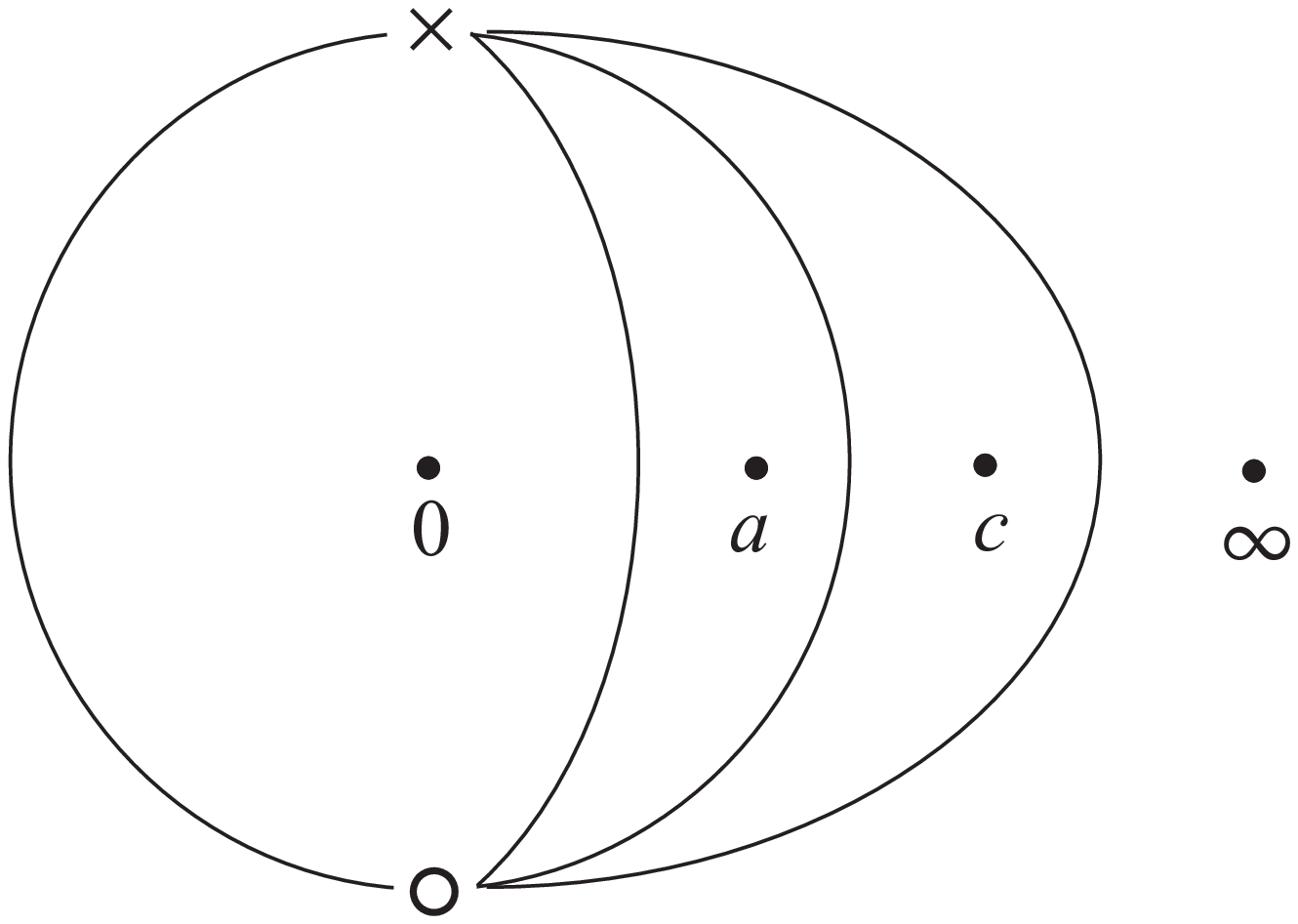}}
\nopagebreak\medskip
\noindent Fig. 7. Cell decomposition $\Phi$ in case $\RR$.
\end{center}

\bigskip

\begin{center}
\epsfxsize=5.0in%
\centerline{\epsffile{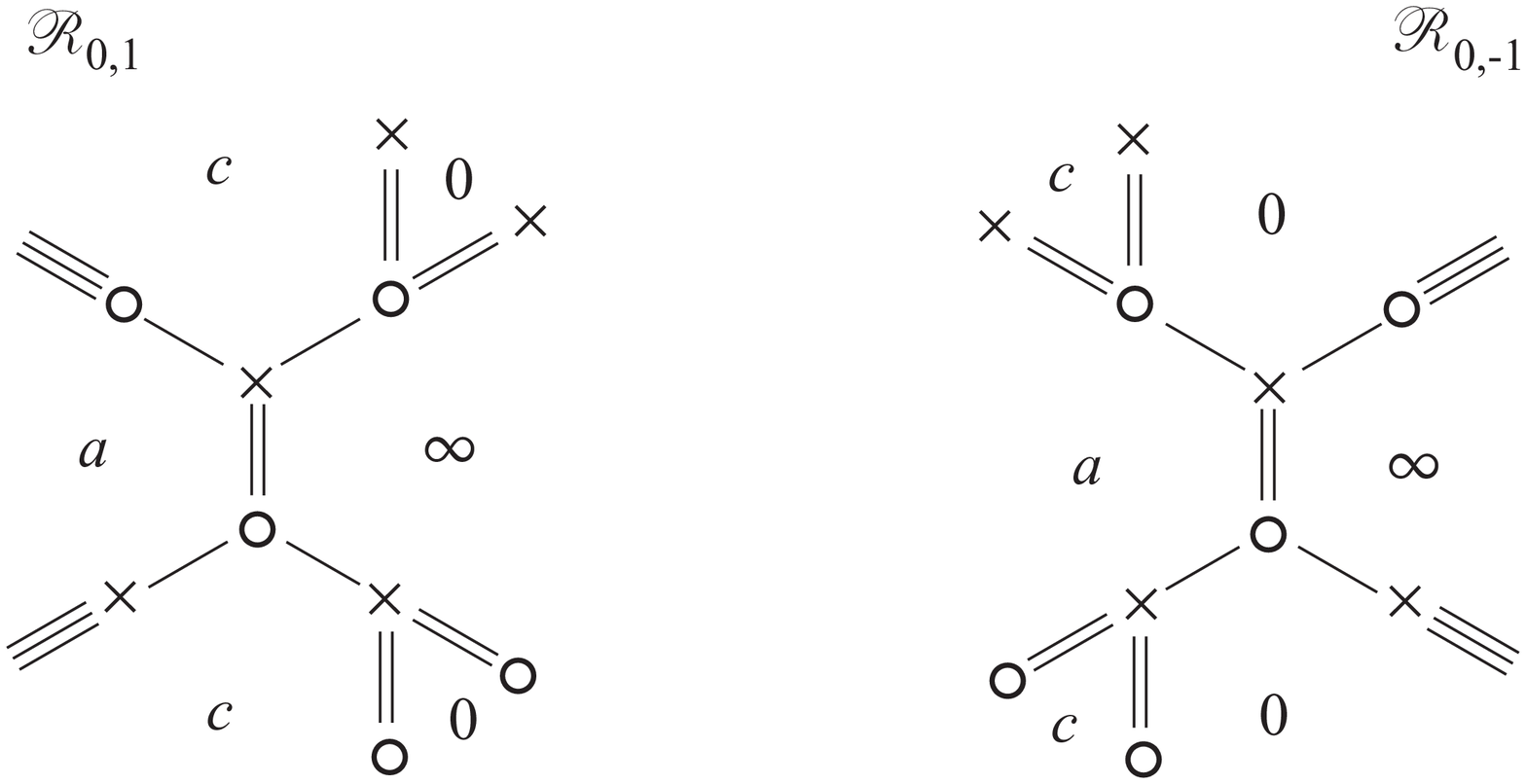}}
\nopagebreak\medskip
\noindent Fig. 8. $\RR_{m,l}$ complexes.
\end{center}
\bigskip

Trees of type $0$ (with $l=0$) are impossible in this case.

Trees $X_{k,l}$ of the type $1$ (with $l>0$) are possible with
$k=2m+1$ and $l$ positive odd. We call the complex $\RR_{m,l}$.
Degeneration $a\to 0$, $t=c/a\to+\infty$ is possible,
and the limit belongs to the QES spectral locus. The limit Nevanlinna
function has $l-1$ zeros, so $J=l$ for this limit function.
Again, using the first part of Theorem 2, we conclude that $J=l$
on the whole image of the Nevanlinna map $F(\RR_{m,l},.)$
Degeneration $c\to+\infty$ (that is $t=c/a\to0+$) gives a Nevanlinna function
for the harmonic oscillator with $J-1+m$ zeros. 
As $\L_{m,l}$ and $\RR_{m,l}$ have common limit of the QES spectral locus,
their Nevanlinna images form a single curve. The results in \cite{EGn} about
QES spectral locus together with counting of zeros of degeneration to
harmonic oscillator show that $\L_{m,l}$ lies on the left and $\RR_{m,l}$ lies
on the right from the intersection point with the QES spectral locus.
See Figs.~11, 13, where Nevanlinna images of $\L_{m,l}$ and $\RR_{m,l}$
are shown with the solid lines, and the QES locus with the dotted line.

Trees $X_{k,l}$ of type $2$ (with $l<0$) give line complexes when $k=2m+1$ and $l$ negative odd.
We call these complexes $\RR_{m,l}$.
Degeneration $c\to\infty$ is possible on the spectral locus.
These charts correspond to the charts $\RR_{m,-l}$ by Darboux transform.

Thus cases $\L$ and $\RR$ and trees of types
1 and 2 cover all cases when $J$ is
odd. We conclude from this: a) $J$ is constant on
the Nevanlinna images of $\L_{m,l}$ and $\RR_{m,l}$, namely $J=l$. 
b) Even values of $J$ must be covered by the remaining trees from
our classification.

{\em Case $\E$.} $\quad 0<c<a$, see Figs.~9--10.
\bigskip

\begin{center}
\epsfxsize=2.0in%
\centerline{\epsffile{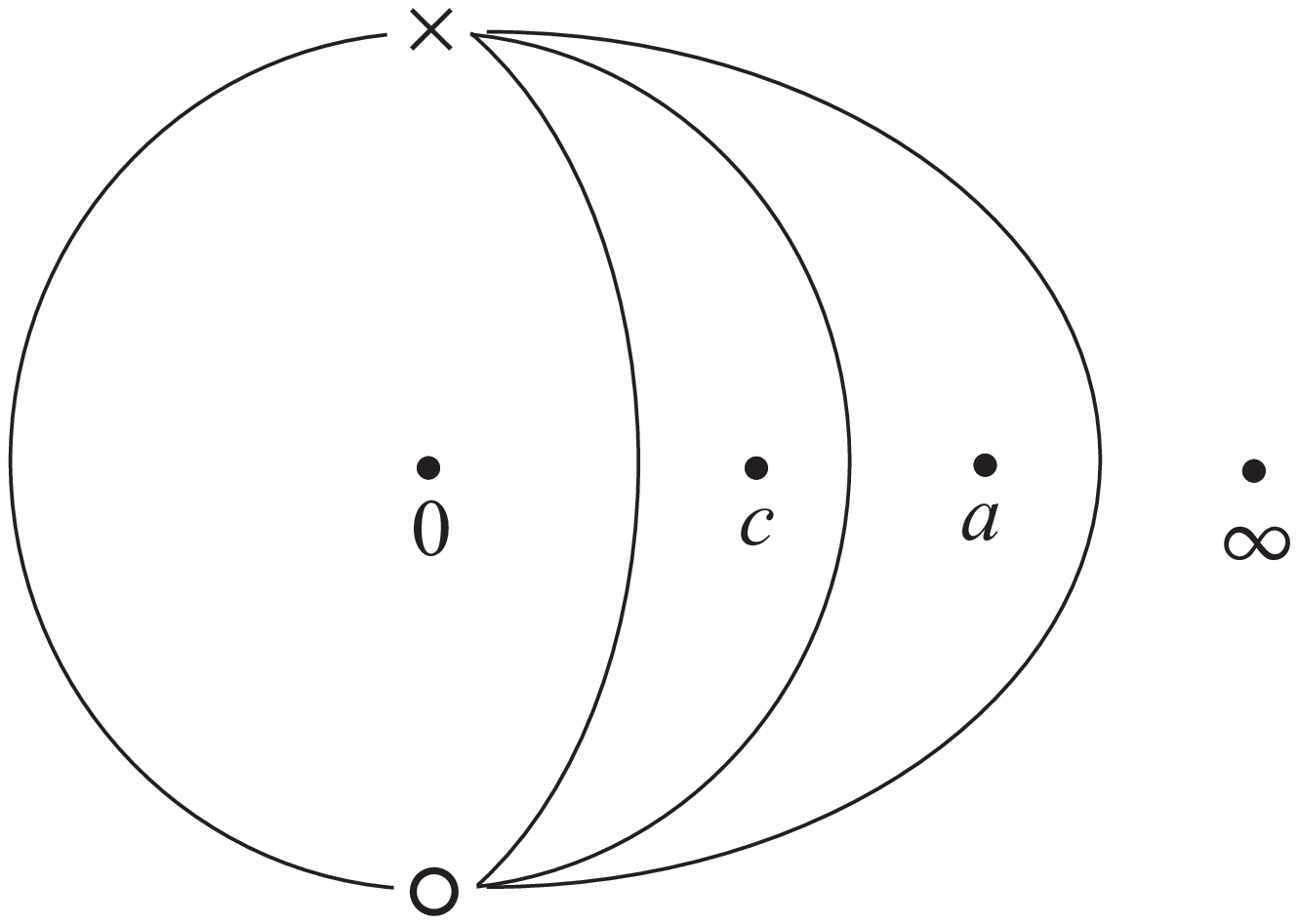}}
\nopagebreak\medskip
\noindent Fig. 9. Cell decomposition $\Phi$ in case $\E$.
\end{center}
\bigskip

\begin{center}
\epsfxsize=5.0in%
\centerline{\epsffile{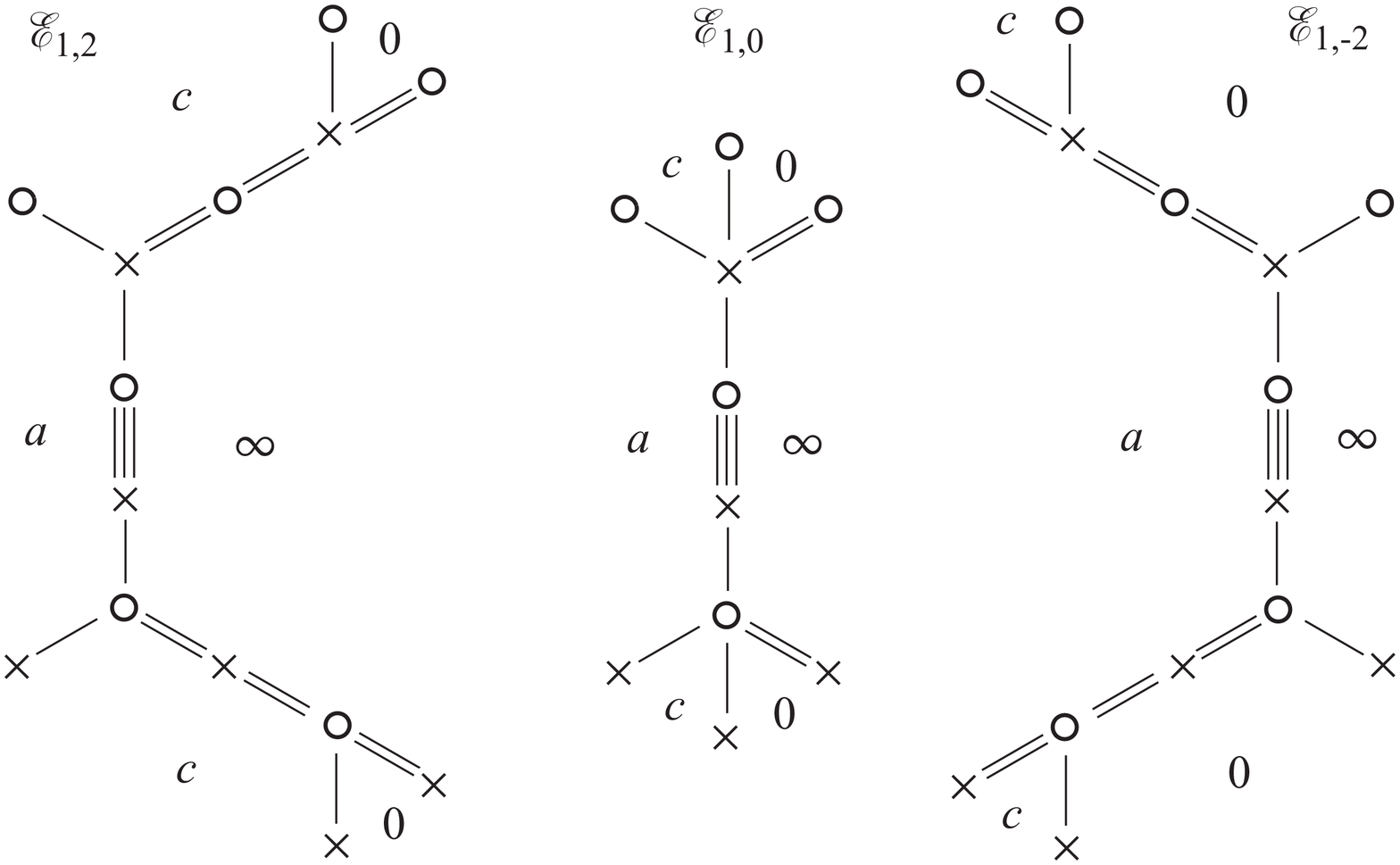}}
\nopagebreak\medskip
\noindent Fig. 10. $\E_{m,l}$ complexes.
\end{center}
\bigskip

Trees $X_{k,0}$ of type $0$ have parameter 
$k=2m+1,\; m\geq 0$, and we denote the corresponding line complex by $\E_{m,0}$.
This line complex represents Nevanlinna functions from section
\ref{J=0} which correspond to $J=0$.

Trees $X_{k,l}$ of type $2$ have $k=2m+1$ and $l$ negative even.
We call the corresponding line complex $\E_{m,l}$. No
degeneration on the spectral locus is possible. We know that for even
$J$ the non-QES spectral locus consists of graphs of functions.
Degeneration as $c\to 0$ and $c\to a$ gives a Nevanlinna function
for the harmonic oscillator with $m$ zeros. Thus these trees parametrize the
whole spectrum for negative even $J$.

Trees $X_{k,l}$ of type $1$ have $k=2m+1$ and $l$ positive even.
We call the corresponding complex $\E_{m,l}$. It corresponds to $\E_{m,-l}$
by the Darboux transform. Degeneration as $c\to 0$ and $c\to a$ gives
a Nevanlinna function
for the harmonic oscillator with $m$ and $m+l$ zeros, respectively.
\vspace{.1in}

These arguments show that real $c$ and $a\neq 0$ correspond to integer
$J$ and that we obtain a parametrization of the whole non-QES spectral locus
in this way. This completes the proof of Theorem 2.

The parametrization of the spectral locus for integer $J$ is represented in
Figs.~11-15. The symbols of line complexes are written below
the corresponding curves. The QES spectral locus is shown with dotted lines.
It was parametrized with different cell decompositions (not with line complexes!)
in \cite{EGn}. Symbols $X_{k,l}$ in the figures refer to the charts
on the QES locus described in \cite{EGn}.   
\bigskip

\begin{center}
\epsfxsize=5.0in%
\centerline{\epsffile{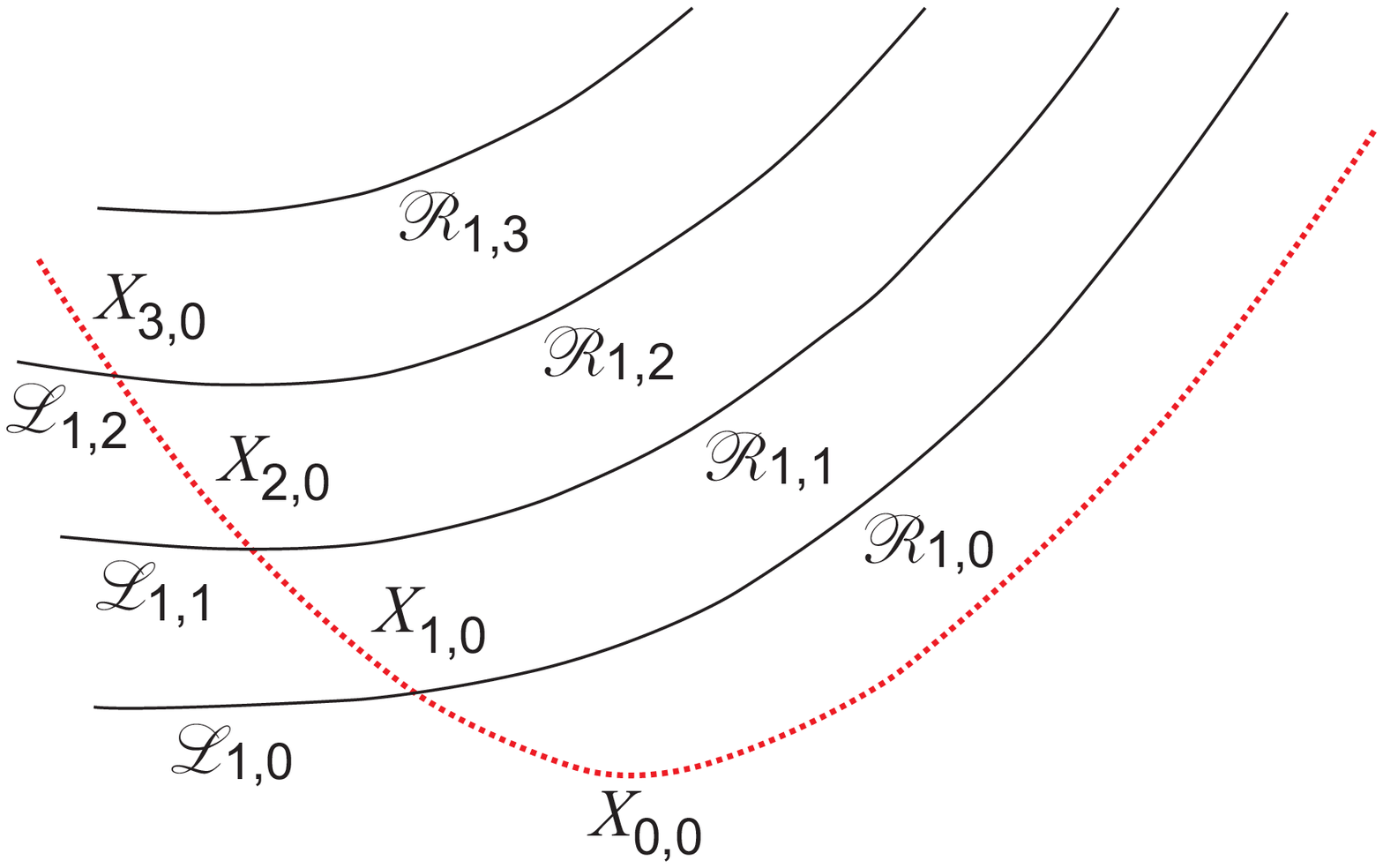}}
\nopagebreak\medskip
\noindent Fig. 11. $Z_{1}(\R).$ 
\end{center}

\bigskip

\begin{center}
\epsfxsize=5.0in%
\centerline{\epsffile{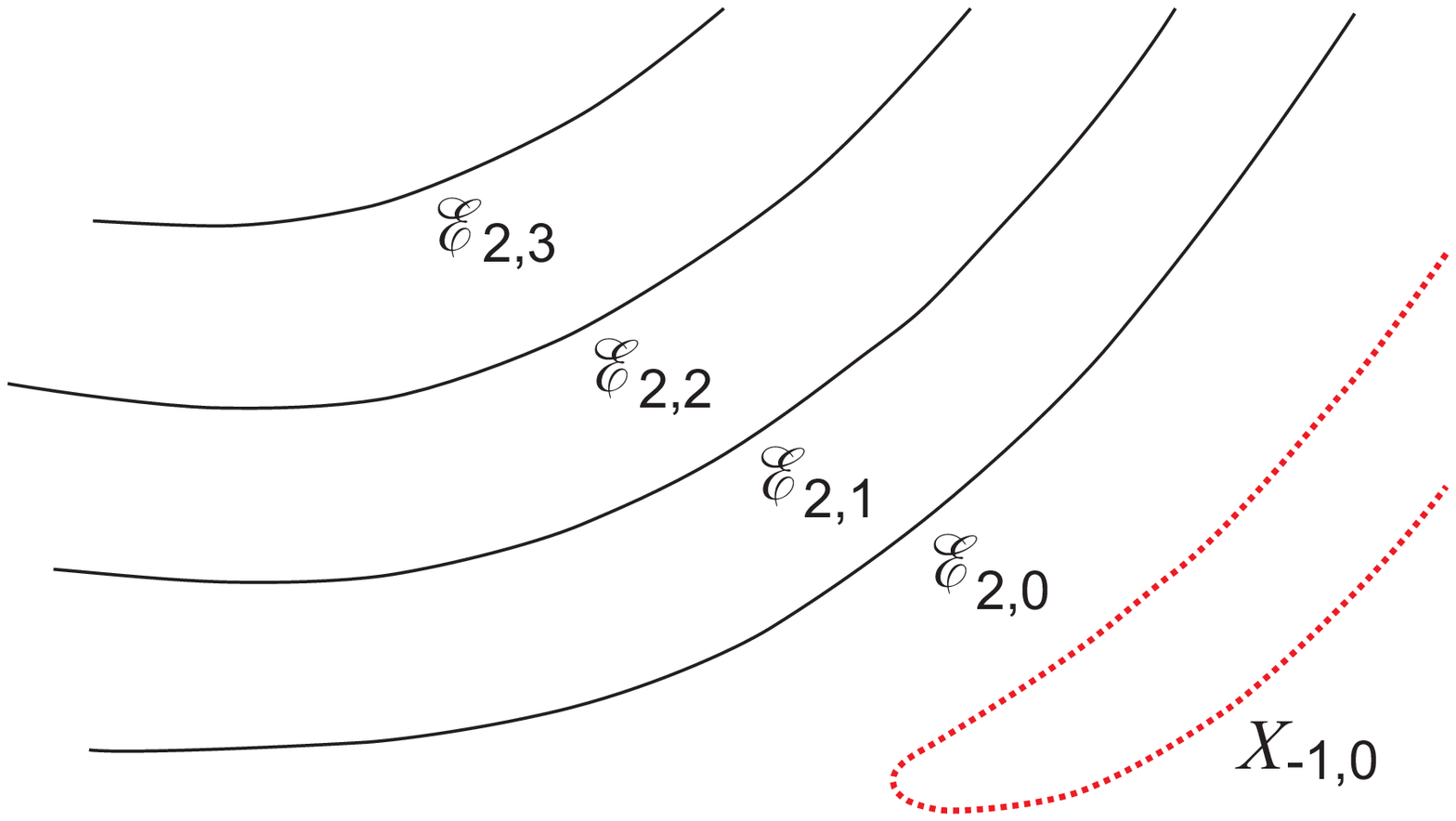}}
\nopagebreak\medskip
\noindent Fig. 12. $Z_2(\R)$.
\end{center}

\bigskip

\begin{center}
\epsfxsize=5.0in%
\centerline{\epsffile{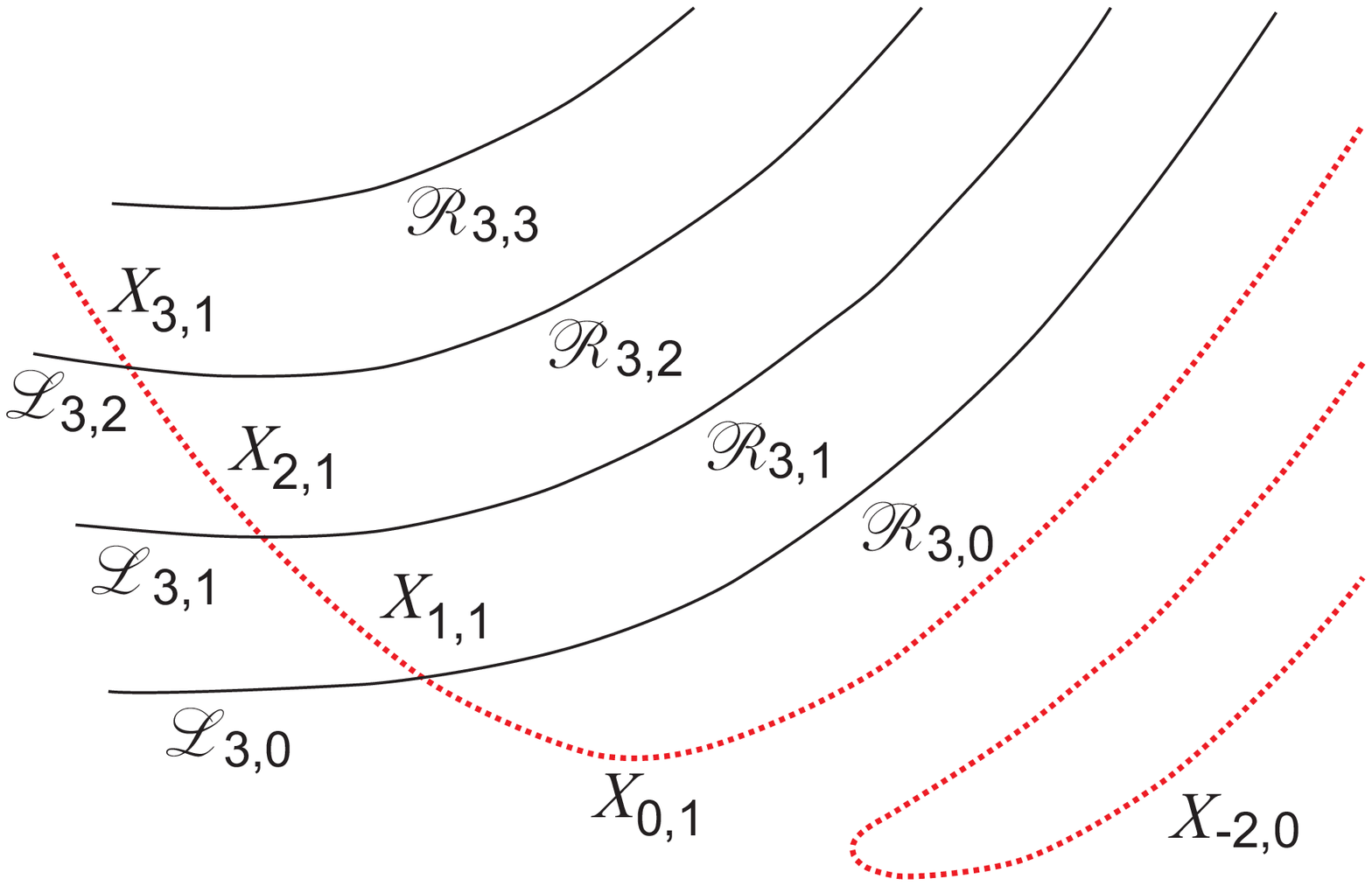}}
\nopagebreak\medskip
\noindent Fig. 13. $Z_3(\R).$
\end{center}

\bigskip

\begin{center}
\epsfxsize=5.0in%
\centerline{\epsffile{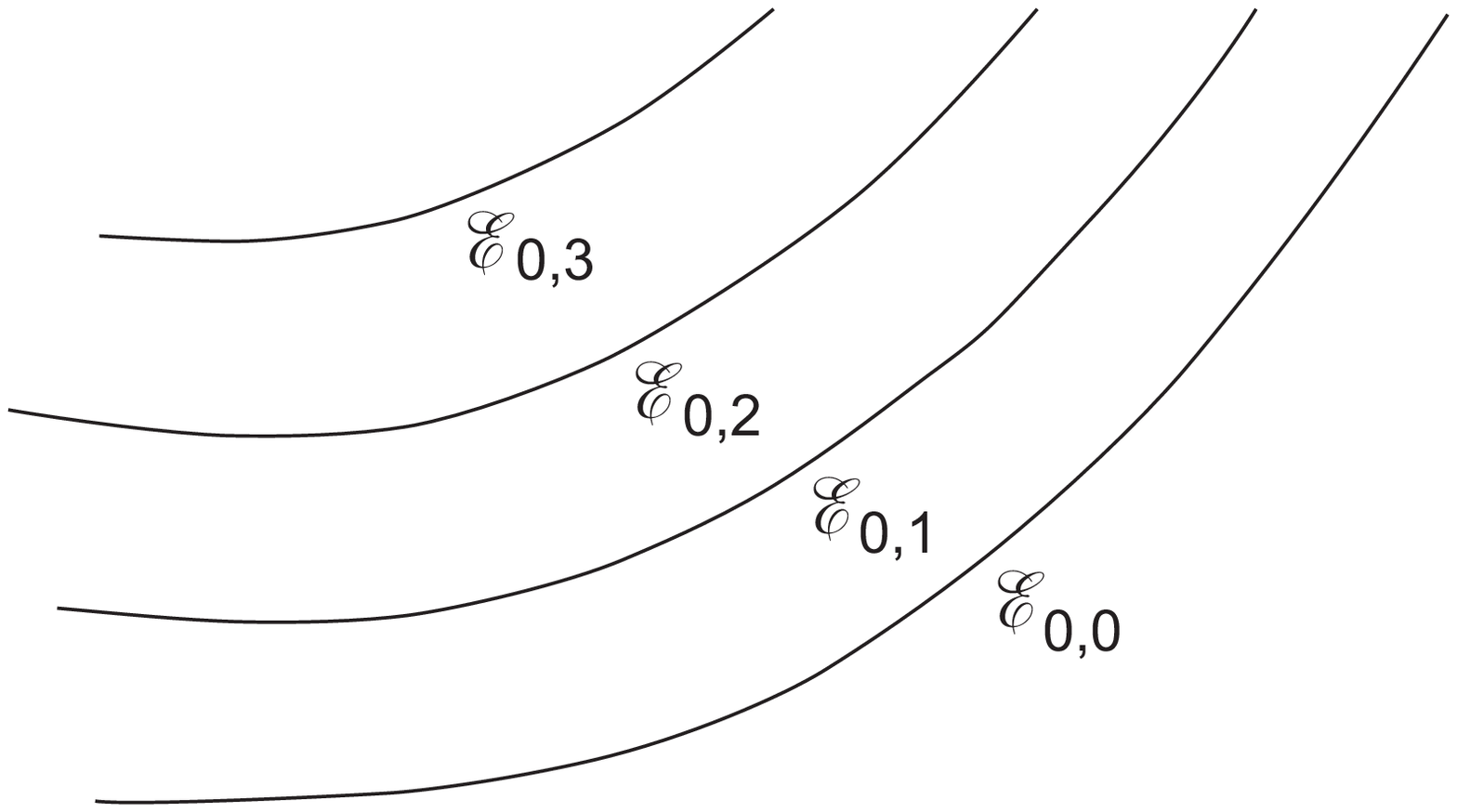}}
\nopagebreak\medskip
\noindent Fig. 14. $Z_0(\R)$.
\end{center}

\bigskip

\begin{center}
\epsfxsize=5.0in%
\centerline{\epsffile{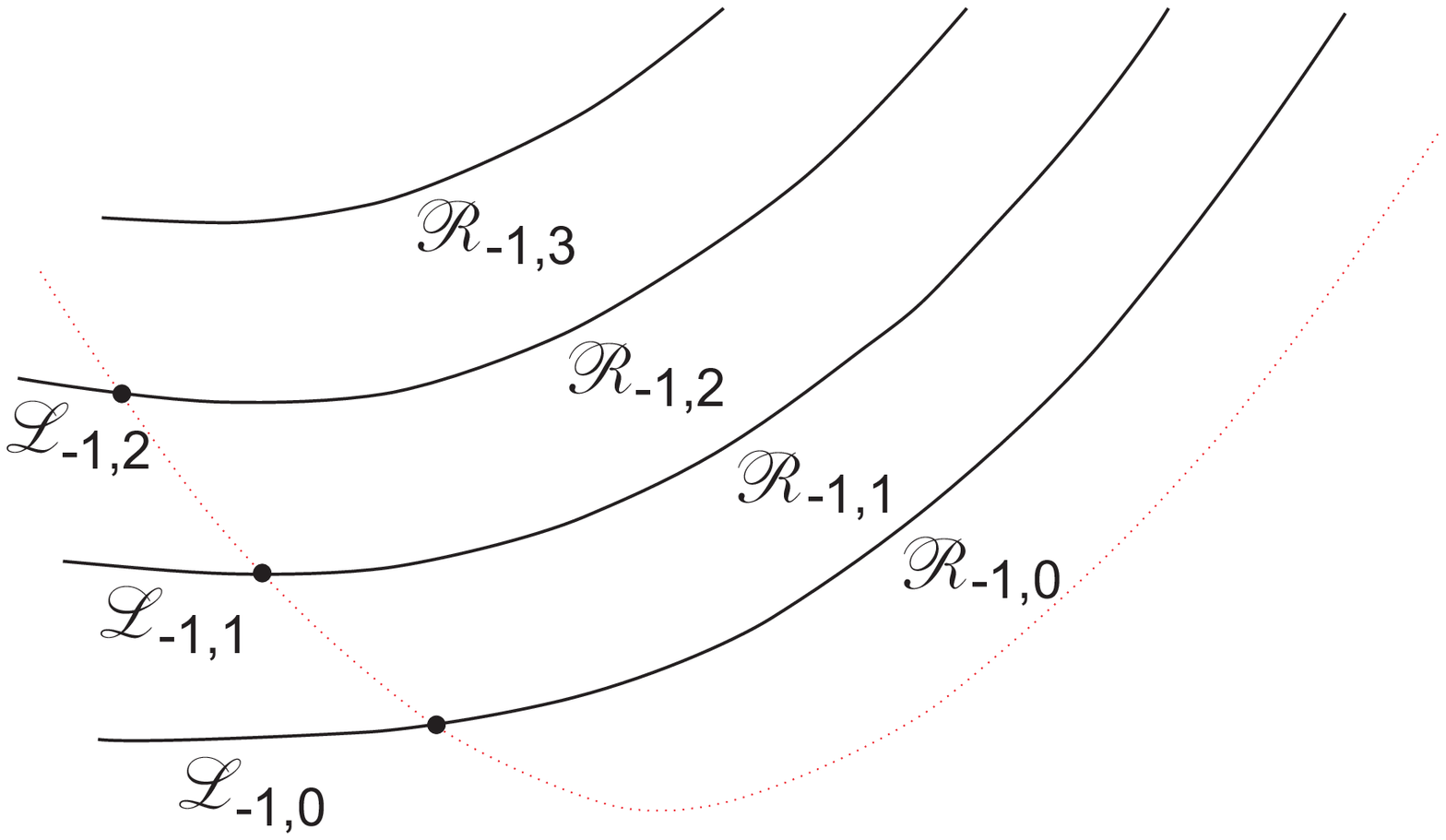}}
\nopagebreak\medskip
\noindent Fig. 15. $Z_{-1}(R)$ superimposed with $Z_1^{QES}(\R)$
(thin dotted line).
\end{center}

{\em Department of mathematics,

Purdue University,

West Lafayette, IN 47907

eremenko@math.purdue.edu

agabriel@math.purdue.edu}

\begin{thebibliography}{1}
\bibitem{Bakken} I. Bakken, A multiparameter eigenvalue problem
in the complex plane,
Amer. J. Math. 99 (1977), no. 5, 1015--1044.
\bibitem{BB} C. Bender and S. Boettcher,  Quasi-exactly solvable quartic potential,
J. Phys. A 31 (1998), no. 14, L273--L277,
arXiv:physics/9801007.
\bibitem{BB2} C. Bender and S. Boettcher, Real spectra in non-Hermitian Hamiltonians
having PT symmetry, Phys. Rev. Lett., 80 (1998)
5243--5246.
\bibitem{Bm} C. Bender, D. Brody, J-H. Chen, H. Jones, K. Milton and C. Ogilvie,
Equivalence of a complex PT-symmetric quartic Hamiltonian and a Hermitian
quartic Hamiltonian with an anomaly, arXiv:hep-th/0605066v2.
\bibitem{BG} V. Buslaev and V. Grecchi, Equivalence of unstable
anharmonic oscillators and double wells, J. Phys. A, 26 (1993) 5541--5549.
\bibitem{Crum} M. Crum, Associated Sturm--Liouville systems,
Quart. J. Math., 6 (1955) 121--127.
\bibitem{DT} 
E. Delabaere, D. T. Trinh, Spectral analysis of the complex cubic oscillator,
J. Phys. A 33 (2000), no. 48, 8771--8796.
\bibitem{Dorey} P. Dorey, C. Dunning and R. Tateo,
The ODE/IM correspondence. J. Phys. A 40 (2007), no. 32, R205--R283.
\bibitem{EG-1} A. Eremenko, A. Gabrielov and B. Shapiro, 
Zeros of eigenfunctions of some anharmonic oscillators, Ann. Inst. Fourier, Grenoble, 58, 2 (2008) 603-624. 
\bibitem{EG1} A. Eremenko and A. Gabrielov, 
Analytic continuation of eigenvalues of a quartic oscillator, Comm. Math. Phys., v. 287, No. 2 (2009) 431-457. 
\bibitem{EG0} A. Eremenko and A. Gabrielov, 
Singular perturbation
of polynomial potentials in the complex domain
with applications to PT-symmetric families, to appear
in Moscow Math. J., arXiv:1005.1696.
\bibitem{EG} A. Eremenko and A. Gabrielov, Quasi-exactly solvable quartic:
elementary integrals and asymptotic, 
J. Phys. A: Math. Theor. 44 (2011) 312001. 
\bibitem{EGn} A. Eremenko and A. Gabrielov, Quasi-exactly solvable quartic:
real algebraic spectral locus, arXiv:1104.4980.
\bibitem{Veselov} J. Gibbons and A. P. Veselov, On the rational monodromy-free
potentials with sextic growth,
J. Math. Phys. 50 (2009), no. 1, 013513, 25~pp.
\bibitem{GO} A. A. Goldberg and I. V. Ostrovskii, Distribution of values
of meromorphic functions, AMS, Providence RI, 2008.
\bibitem{GM} 
V. Grecchi, M. Maioli and A. Martinez,
Pad\'e summability of the cubic oscillator,
J. Math. Phys. A: Math. Theor. 42
(2009) 425208, 17pp.
\bibitem{Masoero} D. Masoero, Y-System and Deformed Thermodynamic Bethe Ansatz,
Lett. Math. Phys. 94 (2010), 151--164.
\bibitem{N} R. Nevanlinna,
\"Uber Riemannsche Fl\"achen mit endich vielen Windungspunkten, Acta Math., 58 (1932) 295--373.
\bibitem{Shin0} K. Shin, Eigenvalues of PT-symmetric oscillators with polynomial potentials,
J. Phys. A 38 (2005), no. 27, 6147--6166.
\bibitem{Shin} K. Shin, On the reality of the eigenvalues for a class of
PT-symmetric oscillators, Comm. Math. Phys. 229 (2002), no. 3, 543--564.
\bibitem{Sibuya} Y. Sibuya,
Global theory of a second order linear ordinary differential
equation with a polynomial coefficient, North-Holland,
Amsterdam; American Elsevier, NY, 1975.
\bibitem{Volk} L. Volkovyski, Converging sequences of Riemann surfaces,
Mat. Sbornik, 23 (65) N3 (1948) 361--382.
\bibitem{Z} J. Zinn-Justin and U. Jentschura, Imaginary cubic perturbation: numerical
and analytic study, J. Phys. A, 43 (2010) 425301.
\end{thebibliography}
\end{document}